\documentclass[prd,preprint,aps,nofootinbib]{revtex4}
\usepackage{color} 
     
\usepackage{amssymb}
\usepackage{amsmath}
\usepackage{graphicx}
     
\usepackage{pdfpages}

\begin{document}

\title{The Impact of Particle Production on Gravitational Baryogenesis}

\author{J. A. S. Lima}
\email{jas.lima@iag.usp.br}
\affiliation{Departamento de Astronomia, Universidade de S\~ao Paulo, Rua do Mat\~ao 1226, 05508-900, S\~ao Paulo, Brazil}
\author{D. Singleton}
\email{dougs@csufresno.edu}
\affiliation{Department of Physics, California State University Fresno, Fresno, CA 93740-8031, USA \\
and \\
ICTP South American Institute for Fundamental Research,
UNESP - Univ. Estadual Paulista
Rua Dr. Bento T. Ferraz 271, 01140-070, S{\~a}o Paulo, SP, Brasil}

\date{\today}

\begin{abstract}
Baryogenesis driven by curvature effects is investigated by taking into account gravitationally induced particle production in the very early Universe. In our scenario, the baryon asymmetry is generated dynamically during an inflationary epoch powered by ultra-relativistic particles. The adiabatic particle production rate provides both the needed negative pressure to accelerate the radiation dominated Universe and a non-zero chemical potential which distinguishes baryons and anti-baryons thereby producing a baryon asymmetry in agreement with the observed value. Reciprocally, the present day asymmetry may be used to determine the inflationary scale at early times. Successful gravitational baryogenesis is dynamically generated for many different choices of the relevant model parameters.   

\end{abstract}
\maketitle

\section{Introduction}

A well known but still challenging  cosmological fact is that the number of baryons in the visible Universe is much larger than the number of anti-baryons.  The baryon asymmetry (B-asymmetry) is usually characterized by the dimensionless quantity:

\begin{equation}
\label{eta}
\eta  = \frac{n_b - n_{\bar b}}{s}\equiv \frac{n_B}{s}\,,
\end{equation}
where $n_{b}$, $n_{\bar b}$ are  the number densities of baryons (anti)-baryons,  respectively, and $s$ is the radiation entropy density. Current constraints on $\eta$  are based on precision measurements of the primordial deuterium abundance combined with the analysis of cosmic background radiation (CMB) acoustic peaks. It lies in the interval $(5.7-6.7) \times 10^{-10}$ \cite{eta}.

The agreement between Big-bang nucleosynthesis predictions and the CMB observations suggests that the above ratio has remained constant at least since the cosmic factory started the production of the light elements.  It is now  widely believed that the B-asymmetry was dynamically generated in the very early Universe  with the $\eta$ value being expressed through some fundamental parameters of particle physics and cosmology.  

Many models have been proposed based on new interactions satisfying (entirely or partially) the well known criteria for baryogenesis advanced in the seminal paper by Sakharov \cite{Sakharov}. However, there is no consensus among cosmologists regarding the correct approach to describe the observed B-asymmetry, nor even whether one needs to strictly adhere to all of Sakharov's conditions (see \cite{review} for discussion of these issues).

In this work, we consider the model dubbed gravitational baryogenesis (GB) which is defined by an effective  derivative  coupling between the Ricci scalar curvature ($R$) and the baryon current \cite{steinhardt}. As in the ``spontaneous baryogenesis" model \cite{spontB}, (which inspired the GB model) the GB approach also leads to an effective chemical potential which is proportional to the time derivative of the Ricci scalar, $\mu \propto \dot R \equiv {dR}/{dt}$, giving rise to a relative shift between the baryon and anti-baryon number. 

Now, for a flat  homogeneous and isotropic FRW  geometry supported by a perfect fluid, the Ricci scalar reads \cite{KT90}:
\begin{equation}
\label{R_FRW}
R\equiv -6 \left( \frac{\ddot a}{a} + \frac{{\dot a}^2}{a^2} \right) = -\left( 1-3\omega \right)\frac{\rho}{M_{Pl}^{2}}\,,
\end{equation}
where $a(t)$ is the scale factor, $\rho$ is the energy density, and $M_{Pl}=(8\pi G_N)^{-1/2} \approx 2.4 \times 10^{18}$ GeV,  is the reduced Planck mass. As usual, the  $\omega$-parameter defines the fluid equation of state (EoS), $\omega = \frac{p}{\rho}=const.$, where $p$ is the pressure.   

From \eqref{R_FRW}, the observed B-asymmetry cannot be generated by the GB mechanism when the cosmic fluid is radiation dominated {\it i.e.} when $w = \frac{1}{3}$. This means the expression in \eqref{R_FRW} must somehow be corrected in order to obtain a non-zero $R$, and $\dot R$ at very early times. 
  
This problem for GB related to the null values of $R$ and $\dot R$ (and thus the vanishing of $\eta$) when the Universe is dominated by ultra-relativistic particles has motivated different solutions in the literature \cite{GS2007,S07,OO2016}. Lambiase and Scarpeta \cite{GS2007} discussed GB in f(R)-gravity theory while Sadjadi \cite{S07} investigated a possible time-variation of $\omega$. More recently, Odintsov and Oikonomou \cite{OO2016}, adopted the Gauss-Bonnet invariant in order to obtain a non-zero $\eta$ even in the radiation domination era (see also \cite{TK04} for a connection with braneworld inspired cosmology and \cite{hook} for GB in context of Hawking radiation from primordial black holes). One aim of this article is to propose a new solution for this problem based on the gravitational particle production in the very early Universe.  
 
There is growing interest in cosmologies driven by gravitationally induced particle production \cite{L1,L2,L3,L4,L5,L6,L7,L8,L9,L12, lima4,LR99}. These papers adopted the non-equilibrium macroscopic description proposed long ago by Prigogine and coworkers \cite{P89} based on the thermodynamics of open systems. A covariant description for the process was advanced in \cite{LCW92}. {It has also been argued that matter creation at the expenses of the gravitational field occurs only as an irreversible process constrained by the usual requirements of non-equilibrium relativistic thermodynamics \cite{P89,LCW92,LG92}. Dynamically, the negative pressure describing matter creation acts like a second viscosity stress, an effective mechanism suggested by Zeldovich \cite{Zeld70} to describe phenomenologically the  cosmic particle production process.   However, it has been proved  that gravitational particle production and bulk viscosity are not equivalent from a thermodynamic viewpoint \cite{LG92}. Although not physically equivalent,  it has been shown that the negative pressure of both  mechanisms  can source inflation (some examples are given in Refs. \cite{Murphy, Barrow86, LPW88, LG92}).}

More recently,  a relativistic kinetic treatment that fully recovers the macroscopic approach for gravitational particle production has also been proposed \cite{L10,L11}. {In principle, this means that an acceptable non-equilibrium theory
for gravitational induced particle production requires finite-temperature quantum field theory
in curved space-times. The lack of such a theory points to a phenomenological
approach in order to incorporate back-reaction in the cosmic dynamics.}

In the macroscopic (or kinetic approach) the back reaction  on  the geometry is included from the very beginning. In particular, the Ricci scalar as given by (\ref{R_FRW})  becomes:

\begin{equation}
\label{RE}
R =-\left( 1-3\omega + (1+\omega)\frac{\Gamma}{H}\right)\frac{\rho}{M_{Pl}^{2}}\,,
\end{equation}
where $\Gamma$, with dimensions of (time)$^{-1}$, is the particle production rate and $H=\frac{\dot a}{a}$ is the Hubble parameter (see section 2 for details). Since $R$ is different from zero for $\omega = \frac{1}{3}$, the extra, phenomenological $\Gamma (H)$ term may potentially produce B-asymmetry even during the radiation phase. Note also that for negligible particle production, $\Gamma (H) \ll H$, the standard result for $R$ is recovered. For an analysis which ignores the effect of the particle production on the Ricci scalar see \cite{modak1,modak2}.

Closely related with the present work is the tepid or warm deflationary model \footnote{Deflationary model here means only an exact but unstable primordial de Sitter state that subsequently deflates towards the standard radiation phase. For a more general definition see \cite{Barrow86}.} driven by gravitationally induced particle production \cite{L2,L5,LR99}. This kind of inflationary scenario is significantly different from isentropic  inflation, as well as from warm inflation \cite{Berera}. Firstly, it is not driven by a scalar field, since its basic mechanism is the gravitational particle production process. Secondly, although filled exclusively by ultrarelativistic particles  ($\omega=\frac{1}{3}$),  its evolution starts from an exact, nonsingular de Sitter state powered by the negative pressure associated with the gravitationally produced thermal bath.  This scenario resembles the idea of a cosmology emerging from nothing, via quantum tunneling, directly into a de Sitter space \cite{VIL82}.  However, different from many variants of inflation, there is no Big-bang singularity (or horizon problem), and the exact, but unstable, primordial de Sitter stage evolves smoothly to the standard radiation FRW phase when the particle production ends -- in agreement with conformal invariance \cite{Parker}. 

In this context, we show that  the observed B-asymmetry is naturally generated during a warm inflationary period with $\omega = \frac{1}{3}$. As we shall see, the proposed solution is not fine-tuned, and by inverting the argument the observed B-asymmetry may also be used to determine the scale of deflation.  

The paper is organized as follows: In section 2 we review briefly how a non-singular de Sitter phase followed by inflation with a ``graceful exit" is naturally powered by adiabatic gravitationally induced particle production. In section 3, we quantify the B-asymmetry predicted by the model. Finally, in section 4, the basic results  are summarized.  

\section{Inflation Induced by Gravitational Particle Production}

In this section we briefly review the inflationary model  powered by ``adiabatic", cosmological particle production, focusing especially on those aspects that will be relevant to  gravitational B-asymmetry, to be discussed in the next section.  

To begin with, let us consider the spacetime described by a flat FRW  geometry 
\begin{equation}
\label{frw-eqn}
ds^2= dt^2 - a^2(t)\left(dx^{2}+ dy^{2} + dz^{2}\right)\,,
\end{equation}
where $a(t)$ is the scale factor. 
In such a background, the Einstein equations and the balance equation for the particle number and entropy density can be written as \cite{L11,LCW92}
\begin{eqnarray}
\rho &=& 3M_{Pl}^{2}\,H^{2}\,, ~\label{feq1} \\
 p+p_c &=& -M_{Pl}^{2}\,\large[2 {\dot H} + 3H^{2}\large]\,,  ~\label{feq2} \\
\dot{n} +3nH &=& n\Gamma\,\longleftrightarrow \frac{\dot N}{N}=\Gamma\,,\label{feq3}\\
\dot{s} +3sH &=& s\Gamma\,\longleftrightarrow \frac{\dot S}{S}=\Gamma\,, 
\label{feq4}
\end{eqnarray}
where $n$ is the particle number density ($N$ is the total comoving number of particles), $s$ is the entropy density ($S$ is the total comoving entropy), and the creation pressure $p_c$ is defined in terms of the creation rate $\Gamma$ by the expression:
 \begin{equation}\label{pc}
 p_c = - (\rho + p){\frac {\Gamma} {3H}}\,,
\end{equation}
while the energy conservation law ($u_{\mu}{T^{\mu\nu}}_{;\nu}=0$) which is also contained in the field equations now becomes \cite{LCW92,L11}:
\begin{equation}\label{ECL}
 \dot\rho + 3H (\rho + p + p_c) = 0\,.
\end{equation}

It should be noticed that the balance equations $\eqref{feq3}$ and $\eqref{feq4}$ imply that $\frac{\dot S}{S}= \frac{\dot N}{N}$ so that the specific entropy (per particle), $\sigma = S/N=s/n$, is conserved.  This condition defines what is meant by ``adiabatic" particle production \cite{LCW92}. Its major implication is that some equilibrium relations, together the general form of the kinetic phase space density, are preserved \cite{L10}. 

In what follows we consider that the early universe is radiation dominated ($\omega=\frac{1}{3}$). In this case, it has been demonstrated \cite{Lima96,L10} that under ``adiabatic" conditions the quantities $\rho_r$, $n_r$ and $s_r$, as a function of the temperature, scale, respectively,  as  $\rho_r \sim T^{4}$, $n_r \propto s_r \sim T^{3}$ (the same as for the equilibrium relations). However, the temperature law is now determined by the corrected differential equation \cite{L10,Lima96}:
\begin{equation}\label{TL}
\frac{\dot T}{T}=-\frac{\dot a}{a} + \frac{\Gamma}{3}\,.
\end{equation}

On the other hand,  by combining Eqs. \eqref{feq1}, \eqref{feq2}, \eqref{pc} with the radiation EoS, it is readily checked that the evolution equation for the Hubble parameter reads: 

\begin{equation}\label{EM}
 \dot H + {2}H^{2}\left( 1- \frac{\Gamma} {3H} \right)=0\,.
\end{equation}
Note that a de Sitter solution ($\dot H = \dot n=0$) supported by radiation is obtained when $\Gamma =3H$. However, this primordial de Sitter solution is unstable since the evolution of the Universe implies that the ratio $\Gamma/3H$ is a time dependent quantity with the model evolving  to the standard radiation FRW phase.  

{\it How is such a transition described}? The late time suppression of the dimensionless ratio $\Gamma/3H$ suggests that it depends on the Hubble parameter, and, more generally, could be expanded in power series of the form \cite{LR99}: 

\begin{equation}
\label{gammaHa}
\frac{\Gamma}{3H} = \alpha+\beta \left( \frac{H}{H_I} \right) 
+ \gamma \left( \frac{H}{H_I} \right)^{2} +...
\end{equation}
where $\alpha$, $\beta$, $\gamma$ are dimensionless constants and $H_I$ is an arbitrary inflationary scale ($\alpha$ must be very small to guarantee a transition to the standard FRW phase). In order to  simplify matters and discuss analytic results, let us consider a two-parameter, phenomenological particle creation rate \cite{L5}
\begin{equation}
\label{gammaH}
\frac{\Gamma}{3H} = \left( \frac{H}{H_I} \right)^{p},
\end{equation}
where the power index $p$ is a positive  constant. { We stress that expressions for the ratio $\frac{\Gamma}{3H}$ given in \eqref{gammaHa} and \eqref{gammaH} are purely phenomenological.  However, there are
some models where the parameters for the particle production are fixed via physical arguments. For example in \cite{modak1} the production rate is fixed by connecting it to the 
Hawking-like radiation in FRW space-time, along the lines first suggested in \cite{Parker}. In this way one obtains $\frac{\Gamma}{H} \propto H^4$. Here we do not assume any particular physical model for particle production but simply use the phenomenological rate given by \eqref{gammaH}.}

In this case, the equation of motion \eqref{EM} becomes: 
\begin{equation}\label{EM1}
 \dot H + {{2}}H^{2}\left(1- \frac{H^p}{H_I^p}\right)=0\,,
\end{equation}
whose solution reads:
\begin{equation}\label{Ha}
H=\frac{H_{I}}{\left[1+D\,a^{2p}\right]^{1/p}}\,, 
\end{equation}
where $D$ is an integration  constant. This solution describes exactly the idea of deflation with an unstable, primordial de Sitter phase followed by a radiation FRW phase\footnote{{ An initial non-singular and unstable de Sitter stage can be generated not only by gravitational particle production as described above. It appears in non-singular models driven by bulk viscosity \cite{Murphy} and also in running vacuum cosmologies \cite{LBS}. The ubiquity of this solution suggests that exotic initial conditions are not required.}}. This can be seen by looking at the two limiting cases:  For $D\,a^{2p} \ll 1$ we find $H=H_I$ while for $D\,a^{2p} \gg 1$ the solution becomes
\begin{equation}\label{a(t)}
  H=\frac{H_{I}}{D^{1/p} a^{2}}\, \rightarrow\,\, a(t) \propto \sqrt{t}. 
\end{equation}
Therefore, the solution \eqref{Ha}  describes a smooth transition from an early, non-singular Sitter stage to the standard, FRW phase and thus gives a natural, ``graceful" exit from de Sitter to the standard radiation dominated epoch (when the particle production ends). This result can also be checked using the expression for the deceleration parameter:
\begin{equation}\label{HF}
q(H) \equiv -\frac{\ddot a}{aH^{2}}=  1-2 \left( \frac{H}{H_I} \right)^{p}.
\end{equation}
For $H=H_I$ one finds $q=-1$  (de Sitter) while  for $H \ll H_I$ the decelerating parameter approaches  $q=1$ (radiation dominated FRW). As should be expected, inflation ends ({\it i.e.} $\ddot a=0$) before the begin of the FRW phase, that is, when the expansion rate reaches the value $H_{end}=2^{-1/p}H_I$.

The behavior in the thermal sector is also easily established. Once the particle production rate is known, the temperature law \eqref{TL} can readily be integrated (in this connection see Refs. \cite{L10,L5,LR99}). As one may check, it is given by:
\begin{equation}\label{HFT}
T(H)= T_I\left(\frac{H}{H_I}\right)^{\frac{1}{2}},\,\,\,\,\,\,T_I = \left(\frac{270}{\pi^2 g_*}\right)^{1/4}{\sqrt{M_{Pl}H_I}}\,.
\end{equation}
where $g_*=\sum g_i$ counts the total number of relativistic degrees of freedom (d.o.f.) near the still arbitrary inflationary scale $H_I$. This result implies that the temperature at the end of inflation ({\it i.e.} when $H=H_{end}$) is essentially defined by two free parameters ({\it i.e.} $p$ and $H_I$) through the expression,  $T_{end} = 2^{-1/2p}\,T_I$, where $T_I$ depends on $H_I$ as given above.

As remarked before, this formalism naturally incorporates the back reaction effects on the geometry. From  Eqs. \eqref{feq1}, \eqref{pc} and \eqref{ECL}  one may check that the modified Ricci scalar is given by \eqref{RE}. Further, by using Eqs. \eqref{feq1}, \eqref{pc}, \eqref{gammaH} and \eqref{EM1}, we find, for a radiation dominated era, the very simple expression for the Ricci scalar
\begin{equation}
\label{REC}
R =- {4}\left( \frac{H}{H_I} \right)^p \frac{\rho}{M_{Pl}^{2}}\,= -{12}H^{2} \left( \frac{H}{H_I}\right)^{p}\,.
\end{equation}
This reduces to the well known de Sitter result for $H=H_I$. As we shall see, the above formula will be crucial for the gravitational baryogenesis process as discussed next. 

\section{Particle Production and Curvature Baryogenesis}

In gravitational particle production models an equal number of effectively massless particles and anti-particles are created \cite{L12}, and thus one would  expect that such models are not useful for baryogenesis. However, in the GB approach, the asymmetry is generated by a derivative coupling between the Ricci curvature scalar and the baryon current $J^{\mu}_{B}$ (or to the baryon $-$ lepton current, $J^{\mu}_{B-L}$). Following the arguments similar to Ref. \cite{steinhardt} we will show that an observationally acceptable B-asymmetry is possible during the radiation phase by virtue of the particle production process discussed in the previous section.  The Lagrangian density for this interaction  takes the form  \cite{steinhardt}
\begin{equation}\label{Int}
\mathcal{L}_{eff}=\frac{1}{M^2_*}(\partial_\mu R) J^{\mu}_{B}\,,
\end{equation} 
where $M_*$ is an unknown  cut-off mass scale of the theory, usually assumed  to be the reduced Planck mass \footnote{The GB model is essentially a gravitational version of ``spontaneous" baryogenesis approach based on the coupling between $J^{\mu}_{B}$ and the four-gradient of a scalar field, $\frac{1}{f} \partial_\mu \phi J^\mu _{B}$, where $f$  is a cut-off in the effective field theory  \cite{spontB}.}. Such an interaction term  can be obtained from a low-energy quantum gravity approach, as well as in higher dimensional supergravity theories  \cite{steinhardt,KU85}. { It is this interaction term
in \eqref{Int} which is the source of the B violation.}

In the FRW spacetime all physical quantities vary only temporally, hence one may replace $\partial _\mu  R \rightarrow {\dot R}$,  and using \eqref{Int}, one can define an effective  chemical potential, $\mu(t)n_{B} \equiv \mathcal{L}_{eff} = \frac{1}{M^2 _*} (\partial _0 R ) J^{0}_B$. For a species of particle, $i$, carrying a baryon charge, $q_i$, the chemical potential is given by
\begin{equation}
\label{chem-pot}
\mu _i = q_i\frac{{\dot R}}{M^2 _*} = 
\pm \frac{{\dot R}}{M^2 _*} ~,
\end{equation}
where for simplicity we have assumed in the last step that all baryons have baryon number $+1$ and all anti-baryons have baryon number $-1$. In principle, one could also consider cases where particles might carry fractional baryon number or higher integer baryon number, however, the basic conclusions are not changed significantly. { It is this chemical potential, which derives from the effective Lagrangian in \eqref{chem-pot}, that leads to the B-asymmetry. Note as ${\dot R} \rightarrow 0$ that $\mu _i \rightarrow 0$ which implies that the B-violation turns off. As we will see later, after inflation ${\dot R}$ rapidly goes to 
zero so that B-asymmetry generation also rapidly turns off after inflation.} With these assumptions, the B-asymmetry produced by the above chemical potential reads \cite{steinhardt}:
\begin{equation}
\label{eta-2}
\eta = \frac{n_B}{s} \approx \frac{{\dot R}}{M_* ^2 T} {\Big \rvert} _{T=T_D}\,,
\end{equation}
with $T$ evaluated at the temperature $T_D$ when the B-violation operator decouples. In general, the value of $T_D$ is fine-tuned to occur at a definite moment in order to get the desired value of $\eta$. In order to avoid the $\eta$-dilution during  the inflationary process,  $T_D$ is usually identified with the temperature at the end of inflation. Here, by exploring two different  possibilities, we show the robustness of the prediction of $\eta$ in the present scenario.   

In point of fact, the above approximate result \eqref{eta-2}  can be rigorously justified by observing that the radiation entropy per unit volume reads
\begin{equation}
\label{entropy}
s = \frac{\rho + p }{T}=\frac{2 \pi^2}{45} g_{*s} T ^3  ~,
\end{equation}
where $g_{*s}$ is also a sum over the relativistic d.o.f. similar to $g_*$. From now on we assume that all particles are in the ultra-relativistic regime with a common temperature so that $g_{*s} = g_*$. In addition, for a single baryon species the concentration $n_B$  can be calculated by integrating the Fermi-Dirac distributions for baryons and anti-baryons by taking into account the different chemical potentials
\begin{equation}
\label{delta-N}
n_B = \int \frac{d^3p}{(2 \pi )^3} 
\frac{1}{e^{(E  - \mu)/ T} +1}
- \int \frac{d^3p}{(2 \pi )^3} \frac{1}{e^{(E + \mu)/ T} +1} ~.
\end{equation}
Note that the chemical potentials in the above expression appear differently for baryons and anti-baryons since they have opposite baryon numbers, $\pm 1$. It is this difference in chemical potential which is responsible for generating the B-asymmetry. In \eqref{delta-N} we have used the Fermi-Dirac distribution exclusively, since in the Standard Model only fermions carry baryon number. If one assumed that bosons could carry baryon number then one should also use the Bose-Einstein distribution. Since the expression in \eqref{delta-N} is only for a single species of baryon/anti-baryon, one needs to sum over all baryonic degrees of freedom, ({\it i.e.} 
$g_{*b} \equiv \sum _{i= baryons} g_i$). Thus to get the full result for all the baryons one should multiply $n_B$ from \eqref{delta-N} by $g_{*b}$. 

Now, by taking the limit $\mu \ll T$, one can integrate \eqref{delta-N} and using  \eqref{chem-pot} one obtains:
\begin{equation}
\label{delta-N2}
n_B \approx g_{*b}\left( \frac{\mu ^3}{6 \pi^2  } + 
\frac{\mu  T^2}{6 } \right) \approx g_{*b} \frac{\mu  T^2}{6 } \approx \frac{g_{*b}}{6}\frac{{\dot R} T^2}{M_* ^2} \longrightarrow \eta\approx\frac{\dot R}{M_*^{2}T}\,.
\end{equation}
In the above expression two different approximations were made.  First, we have dropped $\mu^3$ relative to $\mu T^2$ again using $\mu \ll T$. Second, in the last step, we have taken $\frac{15g_{*b}}{4 \pi^2 g_*}$ of order unity.  Actually, $g_* > g_{*b}$ but we assume that the difference will not be more than one order of magnitude.  Note that the B-asymmetry parameter in \eqref{eta-2} is determined by $T=T_D$ and ${\dot R}$. As discussed in the introduction, the back reaction of the particle production process implies that R is different from zero even for $\omega=\frac{1}{3}$. One of the advantages of using the GB mechanism in conjunction with the particle creation mechanism is that the particle creation itself modifies $R$ and ${\dot R}$ so that at tree level one can have baryogenesis without resorting to higher order loop calculations to deal with the problem at $\omega =\frac{1}{3}$, as was done in \cite{steinhardt}.  

In order to obtain the expression for the B-asymmetry in the presence of gravitational particle production, we need to calculate ${\dot R}$, which we do by  differentiating \eqref{REC} to give
\begin{equation}
\label{t-dot-R-1}
{\dot R} = -4 \frac{\dot \rho}{M_{Pl} ^2} \left( \frac{H}{H_I} \right) ^p
-4 p \frac{\rho}{M_{Pl} ^2} \frac{H ^{p-1}}{H_I ^p} {\dot H} \,
\end{equation}
Using ${\dot \rho} = -3H( \rho + p + p_c )$ from
\eqref{ECL}, and  $\dot H = -  {{2}}H^{2}\left(1- \frac{H^p}{H_I^p}\right)$ from \eqref{Ha}, we find that \eqref{t-dot-R-1} becomes
\begin{equation}
\label{t-dot-R-2}
{\dot R} = 24(2+p)H_I^3\left(\frac{H}{H_I}\right)^{p+3} 
\left[ 1- \frac{H^p}{H_I ^p} \right]\,.
\end{equation}
{ From the above equation one can see that at early times ({\it i.e.} when $H=H_I$) that ${\dot R}=0$ so from \eqref{chem-pot} the chemical potential is zero and there is no B-asymmetry production.
For late times the Hubble parameter decreases so that $H \ll H_I$ and ${\dot R} \rightarrow 0$. This again drives the chemical potential to zero and thus for late times the B-asymmetry production
turns off. It is only in a narrow range between early and late times that ${\dot R} \ne 0$ and B-asymmetry production occurs as we will discuss shortly.} 

Now, inserting \eqref{t-dot-R-2} into \eqref{eta-2} we obtain:
\begin{equation}\label{etaD1}
\eta \approx \frac{24(2+p)H_I^3}{M_*^{2}T_D}\left(\frac{H}{H_I}\right)^{p+3} 
\left[ 1- \frac{H^p}{H_I ^p} \right]\,,
\end{equation}
where we still need to fix the decoupling temperature. We now consider that the B-violation operator decouples at the end of inflation when $H_{end}=2^{-1/p}\,H_I$ and  $T_D\equiv T_{end} = 2^{-1/2p}\,T_I$ (see the discussion below \eqref{HF} and \eqref{HFT}). In this case, the baryogenesis $\eta$-parameter takes the simple form:
\begin{equation}\label{etaD}
\eta \approx 6\frac{(2+p){2^{-5/2p}}H_I^3}{M_*^{2}\sqrt{M_{pl}H_I}}\left(\frac{\pi^{2}g_*}{270}\right)^{1/4} \approx 6(2+p){2^{\frac{p-5}{2p}}} \left(\frac{M_{Pl}}{M_*}\right)^{2}  \left(\frac{H_I}{M_{Pl}}\right)^{5/2},
\end{equation}
where for  $g_* \approx 106$ we have approximated  $(\pi^2 g_*/270)^{1/4} \sim \sqrt 2$. The above result is the main prediction of our work, and its consequences will now be carefully examined. 

To begin with,  we observe that \eqref{etaD1} implies $\eta \approx 0$ for $H=H_I$ (primordial de Sitter stage) and for $H \ll H_I$ (standard FRW radiation phase). It thus follows that  baryogenesis must occur at some moment between the early de Sitter stage and the begin of the standard radiation FRW phase. 

Note also that once $T_D$ had been fixed, the $\eta$ value depends on 3 free parameters:  the power index $p>0$,  which determines the rapidity to end of inflation, ({\it i.e.} when  $\ddot a=0$ -- see the discussion below \eqref{HF}), and the two ratios, $M_*/M_{Pl}$ and  $H_I/M_{Pl}$. The $\eta$ value is weakly dependent on $p$ but varies appreciably with the two ratios of scales.  

It is worth noticing that the fractional variation of temperature between the de Sitter-phase and the end of inflation, $\frac{\Delta T}{T_I}= (T_I-T_{end})/T_I= 1-2^{-1/2p}$,  is relatively short, especially as the index $p$ increases . This means that the $T_D$ does not change appreciably in the corresponding interval, and its value can be chosen fairly broadly -- without fine-tuning -- on the interval where baryogenesis is physically allowed. Following the tradition for adiabatic inflation, we first make the choice $T_D=T_{end}$. 
{Again due to the smallness of $\frac{\Delta T}{T_I}$, the decoupling temperature, $T_D$, can be chosen anywhere in the interval $(T_I , T_{end})$ without greatly altering our basic conclusions.} Of course, this is possible in models with gravitational particle production, but not for adiabatic, inflationary models  driven by scalar fields (in this connection see \cite{RHB03} for baryogenesis in the framework of warm inflationary models). { One might still argue, that even though we have a degree of freedom in choosing $T_D$ in the interval $(T_I , T_{end})$, there is still some degree of fine tuning due to the derived thermodynamic relation $T_I (H_I)$ [see Eq. (\ref{HFT})].  In Table I below we show that this is not the case by obtaining reasonable $\eta$ for a broad range of $H_I$.}

We now put numbers into \eqref{etaD} to illustrate that our model gives values for $\eta$ which agree with the observed value. Let us for example take $p = 1$, and also take the natural choice for the GB scale, $M_*= M_{Pl}$.  This implies from \eqref{etaD} that $\eta \approx \frac{9}{2} (H_I/M_{pl})^{5/2}$. Hence, by assuming that the inflationary scale is $H_I \approx 10^{-4}M_{Pl}$ (in agreement with some analysis), we obtain $\eta \approx 4.5 \times 10^{-10}$, in rough  accordance with the present observations (see the constraints below \eqref{eta}). Reciprocally, given the observed value of the $\eta$ parameter, we obtain a very reasonable value for $H_I \approx 10^{15}$ GeV, the energy scale of the primordial de Sitter stage.

Naturally, such predictions depend on the values assigned to the three free parameters, $p$, $M_*/M_{Pl}$, and $H_I/M_{Pl}$ (as explained before, in the present scenario, $\eta$ is weakly dependent on the value of $T_D$ in its allowed range).  Thus it is interesting to discuss briefly the robustness of the present scenario to give reasonable values for $H_I$ and $\eta$ without the need for fine-tuning.  
\vskip 0.3cm
\begin{table}[h!]
\centering
\begin{tabular}{|c|c|c|c|c|}
\hline  $p > 0$   & \ $M_{Pl}/M_* \geq 1$ & \ $H_I/M_{Pl}\leq 1$             \ & \ $\eta$    
\\   
\hline  $0.07$ & \ $3.0 \times 10^{6}$ & \ $1.0 \times 10^{-5}$ \ & \ $8.8 \times 10^{-10}$         
\\  
\hline  $0.05$ & \ $3.0 \times 10^{8}$ & \ $9.0 \times 10^{-6}$ \ & \ $3.4 \times 10^{-10}$  
\\
\hline  $0.1$ & $5.0\times 10^{4}$ & $1.0\times 10^{-5}$   \ &  \ $4.2 \times 10^{-10}$ 
\\  
\hline $1.0$ & \ $20.0$ & \ $9.0 \times 10^{-6}$  & \     $4.4 \times 10^{-10}$ 
\\  
\hline $1.0$ & $8.0$  & \ $2.0\times 10^{-5}$ & \ $5.2 \times 10^{-10}$ 
\\
\hline $1.0$ & \ $60.0$ & \ $4.5 \times 10^{-6}$  & \ $7.0 \times 10^{-10}$ 
\\
\hline $2.0$ & \ $10.0$ & \ $1.0 \times 10^{-5}$  & \     $4.5 \times 10^{-10}$ 
\\
\hline $3.0$ & \ $100.0$ & \ $1.5 \times 10^{-6}$  & \ $6.6 \times 10^{-10}$ 
\\
\hline $10.0$ & \ $7.0 $ & \ $7.0\times 10^{-6}$  & \ 
$5.4 \times 10^{-10}$ 
\\
\hline
\end{tabular}
\caption{Baryogenesis predictions for $\eta$ for some selected values of the free parameters.}
\end{table}
In Table I, we display the predictions of the B-asymmetry parameter for a large set of selected  values of the free parameters with $T_D=T_{end}$. The values were chosen to give an idea of  how the observed B-asymmetry can be generated by different combinations of the parameters. One can see that for any value of the power index, $p$, it is possible to obtain $\eta$ in rough agreement with observations for reasonable values of the ratios $M_{Pl}/M_*$, and $H_I/M_{Pl}$. From Table I we see that $M_*$ does not need to be equal to the Planck mass in order to obtain the observed B-asymmetry. More interestingly, although not determined like in inflationary models driven by scalar fields, here the primordial de Sitter scale, $H_I$, can be orders of magnitude smaller than the Planck mass. { Finally we note that the values in Table I are
consistent with the bound from \cite{planck} namely $H_I/M_{Pl} < 3.6 \times 10^{-5}$ at 95 \% confidence level.} 

A possible conclusion from Table I is that the prediction of the $\eta$ parameter in this model is rather robust. However, one may worry that the choice of $T_D=T_{end}$, in the short allowed interval for $T$ where baryogenesis may occur, could still represent a moderate fine-tuning. In order to show this is not the case, a different, more realistic choice for the decoupling temperature is now considered. For example, one might more naturally associate $T_D$ with the maximum value of the B-asymmetry production. Using \eqref{etaD1}, and the temperature relationship from \eqref{HFT} to fix $T_D$, we obtain the $\eta$ parameter in the form:
\begin{equation}
\label{eta-3}
\eta \approx 24\sqrt2 (p+2) \left(\frac{M_{Pl}}{M_*}\right)^{2}\left(\frac{H_I}{M_{Pl}}\right)^{5/2}\left(\frac{H}{H_I}\right)^{p + 5/2} 
\left( 1-\frac{H^p}{H_I^p}\right)\,.
\end{equation}
The last two factors  are time dependent and since $H \leq H_I$, both are defined on the same interval [0,1]. However, as the Universe expands and cools, the first term decreases while the second one increases. This means that the baryogenesis $\eta$-parameter has  a maximum value. By 
differentiating \eqref{eta-3} with respect to $H$ one obtains that the maximum occurs at 
\begin{equation}
 {H_*} = {H_I}
 \left( \frac{p+5/2}{2p+5/2} \right)^{1/p}\,,
\end{equation}
and using \eqref{HFT} this leads to a decoupling temperature 
\begin{equation}
\label{t*}
T_*\equiv T_D = T_I\left(\frac{p+5/2}{2p+5/2}\right)^{1/2p}\,.
\end{equation}
Now for $p \gg 1$, the above temperature becomes $T_*\approx T_{end}=2^{-1/2p}\,T_I$ which is exactly the same expression for the temperature at the end of the inflationary process that we previously used for $T_D$ (see the discussion below \eqref{HFT}). In the opposite regime, $p \ll 1$  (but $p$ still greater than zero), one finds that $H_* \rightarrow e^{-2/5} H_I  \approx 0.67 H_I$ and $T_* \rightarrow e^{-1/5} T_I  \approx 0.82 T_I$. If one takes $p = 10^{-3}$, { $M_* = 2.5 \times 10^{-3} M_{Pl}$ and $H_I \sim 1.0 \times 10^{-5} M_{Pl}$ one finds that $\eta \sim 5.1 \times 10^{-10}$}, again in rough agreement with the observed value of $\eta$. Thus even for very small $p$ acceptable values of $\eta$ can be obtained. 

Summarizing, for large and small values of $p$, acceptable values of $\eta$ are obtained using different definitions for the decoupling temperature. In other words, our results for $\eta$ are insensitive to the choice of $T_D$, thereby showing that there is no fine-tuning (not even moderate fine-tuning), provided that the  phenomenological particle production rate is given by \eqref{gammaH}. 

\section{Summary and Conclusions}

In this paper we have investigated the early generation of B-asymmetry driven by curvature effects, in the context of gravitationally induced particle production models. In the relativistic, inflationary scenario adopted here, the early universe is always dominated by ultrarelativistic particles. Inflation is powered by the negative pressure of the gravitational particle production, and its evolution starts from a nonsingular de Sitter phase and deflates  to the standard radiation phase. The key point is that the back reaction of the created particles allows the gravitational baryogenesis process to work properly  before  the beginning of the standard radiation phase when the particle production ends.

In Table I, one may see how the observed baryogenesis depends on the relevant parameters of the model. Based on two different arguments for the decoupling temperature, we have also shown that successful gravitational baryogenesis (without fine-tuning) may happen in this framework. 

We also stress that gravitational  baryogenesis in the presence of particle production does not require new ingredients, like high order loop corrections,  in order to avoid having $\eta =0$ when $\omega = \frac{1}{3}$ as happened in \cite{steinhardt}. In addition, as can be seen in Table I, the cut-off scale of the gravitational baryogenesis, $M_*$, does not need to be equal to the Planck mass in order to generate the observed value  $ \frac{n_B}{s} \sim 10^{-10}$. More interestingly, this value may also be generically obtained for a primordial de Sitter scale, $H_I \sim 10^{-5}M_{pl}$, which is of the order of the GUT scale (see Table I).   

\vskip 0.5cm
{\par\noindent {\bf Acknowledgments:}} JASL is partially supported by CNPq, CAPES (PROCAD 2013) and FAPESP (Brazilian Research Agencies). DS is supported by a 2015-2016 Fulbright Scholars 
Grant to Brazil and by grant $\Phi.0755$  in fundamental research in natural sciences 
by the Ministry of Education and Science of Kazakhstan. DS wishes to thank the ICTP-SAIFR 
in S{\~a}o Paulo for it hospitality. \\

\end{document}